\documentclass[aps, prl, twocolumn,showpacs,floatfix, groupedaddress]{revtex4}
\usepackage{bm}
\usepackage{graphicx}
\usepackage{bm}

\begin{document}

\title{Low temperature high magnetic field $^{57}Fe$ M$\ddot{o}$ssbauer study of kinetic arrest in $Hf_{0.77}Ta_{0.23}Fe_2$}

\author{V. Raghavendra Reddy$^a$, R. Rawat$^a$, Ajay Gupta$^a$, Pallab Bag$^a$, V Siruguri$^b$ and P Chaddah$^a$}

\affiliation{$^a$UGC-DAE Consortium for Scientific Research, University Campus, Khandwa Road, Indore-452001, India.}
\affiliation{$^b$UGC-DAE Consortium for Scientific Research Mumbai Centre, R-5 Shed, BARC, Mumbai-400085, India.}

\begin{abstract}
Low temperature high magnetic field $^{57}Fe$ M$\ddot{o}$ssbauer measurements were carried out on inter-metallic $Hf_{0.77}Ta_{0.23}Fe_2$ compound by following novel paths in H-T space. The ferromagnetic (FM) fraction at 5K and zero magnetic field is shown to depend on cooling field i.e., higher the field higher is the FM fraction.  M$\ddot{o}$ssbauer spectra collected in the presence of 4 Tesla magnetic field shows that the antiferromagnetic (AFM) spins cant with respect to the applied magnetic field and hence, contribute to the total bulk magnetization in this compound. The data also show induced magnetic moment even at the 2a site of AFM phase.  M$\ddot{o}$ssbauer spectra collected using CHUF (cooling and heating in un-equal magnetic fields) protocol shows re-entrant transition when sample is cooled in zero field and measured during warming in 4 Tesla showing FM state as the equilibrium state. The present work is the first microscopic experimental evidence for the de-vitrification of kinetically arrested magnetic state.
 
\end{abstract}


\maketitle\section{Introduction}

The compound $HfFe_2$, which belongs to the Laves phase compounds of $RFe_2$-type where R is transition metal (A=Sc, Ti, Nb, Hf and Ta), has been reported to show polymorphism with one of the phases crystallizing in $MgZn_2$ type hexagonal structure (space group $P6_3/mmc$) and the second one in cubic structure of $MgCu_2$ type \cite{Nishi1982, Nishi1983}. The $MgZn_2$ type hexagonal structure, also known as C14 structure in literature contains two crystallographically inequivalent Fe sites.  The structure consists of atomic layers alternating along hexagonal c-axis in the sequence Fe(6h)-R-Fe(2a)-R-Fe(6h) and so on. While the exchange interaction between the Fe(6h) atoms is always ferromagnetic within the same layer, the interaction between the neighboring Fe(6h) layers depends on the type of R atom, and Fe atoms at 2a site. Therefore, depending on the type of R atoms, these compounds exhibit large changes in the magnetic properties with composition / temperature without any change in the crystal structure and hence, have been extensively explored \cite{Nishi1982, Nishi1983, Dely2008, Belo2006}.  For example, the magnetic ground state is reported to change from ferromagnetism (FM) to anti-ferromagnetism (AFM) at a Ta concentration between x=0.2 to 0.3 in $Hf_{1-x}Ta_xFe_2$ \cite{Nishi1983, Rawa2012}, whereas in $Hf_{1-x}Ti_xFe_2$, this transition is reported to take place in the vicinity of x=0.6 \cite{Dely2008}. Irrespective of magnetic ordering, the fraction of Fe occupying 6h and 2a sites is reported to be in the ratio of 3:1 from structural and M$\ddot{o}$ssbauer data as well \cite{Nishi1982, Nishi1983, Dely2008, Belo2006}. Apart from the bulk magnetization measurements, these compounds were also studied by M$\ddot{o}$ssbauer and perturbed angular correlation techniques to measure the hyperfine interactions in these compounds. The Mossbauer studies of Nishihara et al. \cite{Nishi1983}, on $Hf_{1-x}Ta_xFe_2$  (x=0-0.7) showed discontinuous change in internal hyperfine fields ($B_{INT}$) of Fe across first order AFM-FM transition.  Such discontinuous change in $B_{INT}$ values through the FM to AFM transition is also observed in FeRh system and is explained in terms of difference of the polarization of conduction electrons to $B_{INT}$ in the two states \cite{Shir1963}.  The $B_{INT}$ on 6h and 2a sites of Fe in $Hf_{1-x}Ta_xFe_2$ compounds is of comparable value ($\approx$17 Tesla) in FM state and it reduces drastically in AFM state ($\approx$ 10 Tesla for 6h site and zero for 2a site of AFM phase) \cite{Nishi1982, Nishi1983, Dely2008, Belo2006}.  In FM $HfFe_2$, various contributions to the $B_{INT}$ at both Hf and Fe sites are calculated theoretically, which are found to match excellently with the measured values and it is reported that the magnetism in $HfFe_2$ originates mainly from the Fe atomic magnetism \cite{Belo2006}.  However, only few reports are seen in literature dealing with the mechanism of magnetic transition in these compounds and the studies on this aspect are far from complete.  For example, Nishihara et al. \cite{Nishi1983}, explained their results with the model of itinerant electron magnetism and Delyagin et al. \cite{Dely2008}, invoked the concept of magnetic frustration of Fe(2a) sites which is caused by a non-collinearity of the magnetic moment of Fe(6h) sublattice. 

Recently Rawat et al.\cite{Rawa2012}, have reported bulk magnetization measurements on $Hf_{1-x}Ta_xFe_2$ (x=0.225,0.230 and 0.235) compounds and observed anomalous thermo-magnetic irreversibility at low temperatures due to kinetic arrest of the first order AFM-FM transition. They observed that the region of co-existing AFM and FM phases increases in H-T space with Ta concentration and concluded that due to the interplay of kinetic arrest and supercooling one can have tunable fractions of AFM and FM phases for the same temperature and magnetic field values.  It is to be noted that the phenomena of phase coexistence and kinetic arrest are reported in many compounds such as manganites, shape memory alloys, multiferroics, magnetocaloric materials, other intermetalics and oxides \cite{Chad2010, Burk2012, Wu2006, Kuma2006, Bane2006, Chad2008, Dash2010, Bane2009, Chad2008R, Maci2007, Shar2008, Bane2011, Shar2007, Sark2011, Kush2008, Kush2009, Roy2007, Roy2006, Chat2005, Choi2010, Angs2008}.  In compounds such as manganites there is a huge difference between the resistivity of AFM and FM states and therefore, the resistivity measurements in conjunction with bulk magnetization measurements are employed to study such phenomena.  However, from such magnetic and transport data, it is difficult to obtain the accurate quantitative information about the fraction of FM and AFM phases and their individual response to the applied magnetic field.  

In the present work we have carried out $^{57}Fe$ M$\ddot{o}$ssbauer measurements under the application of external magnetic fields to study the phenomena of phase coexistence and kinetic arrest in $Hf_{0.77}Ta_{0.23}Fe_2$ compound. It is to be noted that, because of significant difference in the internal hyperfine field ($B_{INT}$) values at Fe sites corresponding to FM and AFM phase of this compound, one can unambiguously estimate these fractions to support the concept of kinetic arrest.  The results presented here clearly demonstrate the phenomena of kinetic arrest in this compound.  Also, in contrast to bulk magnetic measurements, which essentially measure the response of the sample as a whole to the applied magnetic field, one can estimate the individual response of FM and AFM fractions to the applied magnetic field from the M$\ddot{o}$ssbauer measurements as discussed below.  And finally, because of polarization selection rules with the application of external magnetic fields, the M$\ddot{o}$ssbauer data gives unambiguous information regarding the linearity of the magnetic structure, which is expected to give inputs to explain the magnetic phase transition as proposed by Delyagin et al \cite{Dely2008}. 

\maketitle\section{Experimental Details} 
The studied polycrystalline sample of $Hf_{0.77}Ta_{0.23}Fe_2$ was prepared by arc melting the constituent elements under inert argon gas atmosphere and the details of preparation are reported elsewhere \cite{Rawa2012}. The prepared sample is found to be single phase from powder x-ray diffraction measurements and the refinement of the diffraction pattern shows that the alloy crystallizes in $P6_3/mmc$ hexagonal structure.  The ingots were crushed to powder and mixed with silicone grease for $^{57}Fe$ M$\ddot{o}$ssbauer measurements, which were carried out in transmission mode with $^{57}Co$(Rh) radioactive source in constant acceleration mode using standard PC-based M$\ddot{o}$ssbauer spectrometer equipped with WissEl velocity drive.  Velocity calibration of the spectrometer is done with natural iron absorber at room temperature.  For low temperature high magnetic field measurements, the sample was placed inside a Janis make superconducting magnet and the external magnetic field was applied parallel to the  $\gamma$-ray (i.e., longitudinal geometry).  The M$\ddot{o}$ssbauer data was analyzed using NORMOS SITE \cite{Bran1990} program for the extraction of hyperfine parameters. Magnetization measurements were performed using a 7 Tesla SQUID-VSM of Quantum Design, USA on the same sample (alongwith silicon grease) which is used for M$\ddot{o}$ssbauer measurements.

\maketitle\section{Results and Discussions}

Figure 1 shows the temperature dependent $^{57}Fe$ M$\ddot{o}$ssbauer data at the indicated temperatures.  Spectra recorded at 300K and down to 65K were fitted by considering two Fe sites corresponding to Fe at 6h and 2a sites of AFM phase.  The observed isomer shift values are $-0.11 \pm 0.01$ and $-0.07 \pm 0.01$ mm/s; quadrupole splitting values are $0.11 \pm ±0.01$ and $0.32 \pm 0.02$ mm/s for the 6h and 2a sites, respectively.  These values match very well with the literature values.  Also, as reported in literature, we have not observed any significant variation of these parameters across the FM-AFM transition \cite{Nishi1982, Nishi1983}.   It is the $B_{INT}$, which changes drastically across the FM-AFM transition and hence only this parameter including the linewidth (FWHM), area fraction of different magnetic phases, and area ratio of second and third line intensity ($A_{23}$) are shown in the obtained hyperfine parameters as shown in table-1.  It is to be noted that the estimated area fraction of Fe in 6h and 2a sites is always in the ratio of 3:1, which is consistent with the crystallographic data on these compounds \cite{Nishi1982, Nishi1983, Dely2008, Belo2006}.  The $A_{23}$ parameter of the sextet was observed to be close to two which is expected.  Since the measurements are carried out on powder samples and in zero external magnetic fields, one would expect an isotropic distribution of angles between the incident  $\gamma$-ray and the internal hyperfine field corresponding to the random distribution of Fe spins.  The variation of $B_{INT}$ corresponding to AFM 6h site as a function of temperature is observed to be similar to that of Nishihara et al \cite{Nishi1983}. At 5K, the data is fitted with an overlap of two sextets and a doublet, out of which one magnetic sextet and a central doublet correspond to 6h and 2a sites of an AFM phase, respectively, and the outer sextet with a hyperfine field of about 17 Tesla with 7.5\% area fraction corresponds to the ferromagnetic (FM) phase.

\begin{figure}[htb]
	\begin{center}
	\includegraphics[width=8 cm] {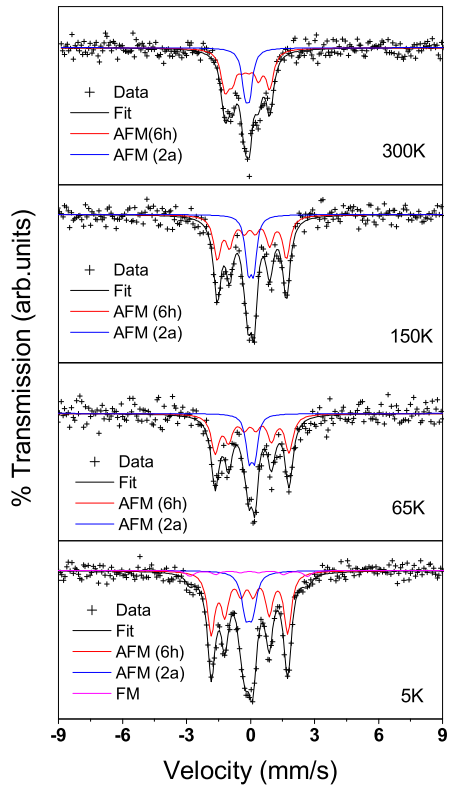}
	\end{center}
	\caption{(color online) Zero field $^{57}Fe$ M$\ddot{o}$ssbauer spectra of $Hf_{0.77}Ta_{0.23}Fe_2$ at the indicated temperatures.  Inset shows the variation of internal hyperfine field ($B_{INT}$) corresponding to the antiferromagnetic Fe(6h) site as a function of temperature. The dots are the experimental points and the solid lines are the best fits to the data. Obtained hyperfine parameters are shown in Table-I.}
	\label{Figure1}
\end{figure}

Figure 2 shows the $^{57}Fe$ M$\ddot{o}$ssbauer spectra measured in zero applied external magnetic fields ($B_{EXT}$) at 5K after cooling in different magnetic fields to study the history dependence of the magnetic states. The sample was cooled from 150K to 5K in different magnetic fields and after reaching 5K, the $B_{EXT}$ is reduced to zero i.e., the measurements are carried out with the sample at 5K in remanent magnetic state, but with different history of field cooling.  The spectra are fitted with two sextets and one doublet by keeping all the parameters as free (except in 5 Tesla remanent data, where the area ratio of AFM 6h and 2a sites is constrained to be equal to 3:1 due to relatively poor statistics). The estimated hyperfine parameters are shown in Table-II. One can clearly see from the data that, as the strength of cooling magnetic field is increased, the FM fraction with $B_{INT}$ of about 17 Tesla increases in intensity.  The FM fraction obtained from these measurements as a function of cooling field at 5K and zero Tesla is shown in figure 3.  

\begin{figure}[htb]
	\begin{center}
	\includegraphics[width=8 cm] {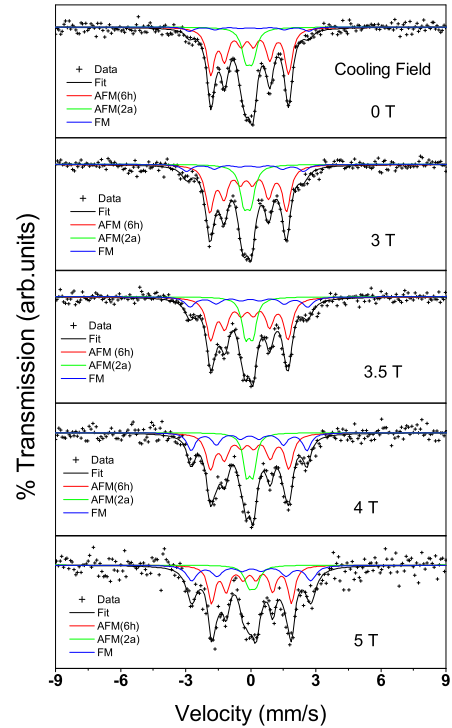}
	\end{center}
	\caption{(color online) Zero field $^{57}Fe$ M$\ddot{o}$ssbauer spectra of $Hf_{0.77}Ta_{0.23}Fe_2$ measured at 5K. Sample was cooled from 150 K in the presence of applied magnetic fields ($B_{EXT}$) as indicated in the figure. The $B_{EXT}$ is reduced to zero isothermally at 5 K and the spectra are collected in remanent state.  The dots are the experimental points and the solid lines are the best fits to the data.  Obtained hyperfine parameters are shown in Table-II.}
	\label{Figure2}
\end{figure}

\begin{table}[htb]
\caption{\label{arttype} Obtained hyperfine parameters from the fitting of $^{57}Fe$ M$\ddot{o}$ssbauer spectra measured at zero field (shown in figure 1) at the indicated temperatures approached by zero field cooling.}
\begin{ruledtabular}
\begin{tabular}{ccccccccc}
Temp & FWHM & $B_{INT}$	& $A_{23}$	& Area & Phase\\
(K) & (mm/s) & (Tesla)	& 	& \% & \\
300 K & 0.451$\pm$0.024 & 6.49$\pm$0.06	  & 2.00	& 75.2 & AFM(6h)\\
      & 0.364$\pm$0.032 & -           	  & -	    & 24.8 & AFM(2a)\\
150 K & 0.426$\pm$0.021 & 10.04$\pm$0.06	& 2.00	& 74.3 & AFM(6h)\\
      & 0.326$\pm$0.035 & -           	  & -	    & 25.7 & AFM(2a)\\
 65 K & 0.451$\pm$0.026 & 10.73$\pm$0.07	& 2.00	& 77.1 & AFM(6h)\\
      & 0.335$\pm$0.048 & -           	  & -	    & 22.9 & AFM(2a)\\
  5 K & 0.460           & 17.04$\pm$0.33	& 2.00	& 7.5  & FM     \\
      & 0.426$\pm$0.014 & 11.12$\pm$0.03	& 2.00	& 68.2 & AFM(6h)\\
      & 0.455$\pm$0.035 & -           	  & -	    & 24.3 & AFM(2a)\\
\end{tabular}
\end{ruledtabular}
\end{table}

\begin{table} [h]
\caption{\label{arttype} Obtained hyperfine parameters from the fitting of $^{57}Fe$ M$\ddot{o}$ssbauer spectra measured at zero field at 5 K approached by cooling from 150 K ((shown in figure 2)) under indicated field values.}
\begin{ruledtabular}
\begin{tabular}{ccccccccc}
Cooling & FWHM & $B_{EFF}$ & $A_{23}$	& Area & Phase\\
field & (mm/s) &(Tesla)	& & \% & \\

3 Tesla   & 0.533$\pm$0.11 & 16.74$\pm$0.24	  & 2.00	& 12.8 & FM    \\
          & 0.463$\pm$0.017 & 10.98$\pm$0.04	& 2.00	& 64.4 & AFM(6h)\\
          & 0.447$\pm$0.036 & -           	  & -	    & 22.8 & AFM(2a)\\
3.5 Tesla & 0.563$\pm$0.078 & 16.84$\pm$0.17	& 2.00	& 20.3 & FM    \\
          & 0.483$\pm$0.021 & 11.05$\pm$0.05	& 2.00	& 60.7 & AFM(6h)\\
          & 0.409$\pm$0.038 & -           	  & -	    & 19.0 & AFM(2a)\\
 4 Tesla  & 0.460$\pm$0.056 & 16.54$\pm$0.13	& 2.00	& 26.3 & FM    \\
          & 0.479$\pm$0.027 & 11.21$\pm$0.07	& 2.00	& 56.2 & AFM(6h)\\
          & 0.364$\pm$0.040 & -           	  & -	    & 17.5 & AFM(2a)\\
  5 Tesla & 0.577$\pm$0.076 & 17.04$\pm$0.17	& 2.00	& 39.2 & FM     \\
          & 0.399$\pm$0.028 & 11.43$\pm$0.09	& 2.00	& 45.7 & AFM(6h)\\
          & 0.466$\pm$0.076 & -           	  & -	    & 15.1 & AFM(2a)\\

\end{tabular}
\end{ruledtabular}
\end{table} 

Using CHUF (cooling and heating in un-equal fields) protocol \cite{Bane2009R} it has been shown that FM state is the ground state in this compound \cite{Rawa2012}.  The M-T data of Rawat et al. \cite{Rawa2012} suggest that cooling in magnetic field higher than 3 Tesla results in close to 100\% FM phase fraction.   Whereas, the FM fraction from the present study is found to be only about ~39\% for cooling in 5 Tesla, which is significantly lower. The powder sample used in the present study was taken from the same ingot which was used in the earlier studies as reported in ref \cite{Rawa2012}. The difference between these two studies could be due to sample mounting conditions for these two measurements. For Mossbauer study, powder sample with silicon grease was used, whereas the earlier magnetization studies were performed on a piece of ingot. To find out the origin of the difference between these two measurements, we performed magnetization measurement on same powder sample (along with grease) which was used for Mossbauer measurements. The magnetization measurement for this sample is shown in figure 4.  Main panel of this figure shows M-H curve measured after zero field cooling and magnetic field is varied from 0 to 7 to 0 to 7 Tesla isothermally at 5 K.  As can be seen from this figure the magnetic field required for AFM to FM transition is distinctly higher than that observed in previous work for same but bulk ingot sample. The second field increasing cycle shows that roughly 50\% of the transformed FM phase is retained down to 0 Tesla. A comparison with saturation magnetization suggests about 8\% FM phase for zero field condition.  To compare the remanant state after field cooling in 4 Tesla M-H curve is measured after field cooling in 4 Tesla from 150 K to 5 K and then M is measured with reducing magnetic field from 4 Tesla to zero and again back to 4 Tesla. It shows that about 50\% FM fraction is obtained in 4 Tesla cooling and the second field increasing cycle shows that most of this phase is retained when magnetic field is reduced to zero. This is consistent with above analysis of Mossbauer data. Since these measurements were performed with a powder sample mixed with silicon grease, the absolute determination of magnetization is not possible. We have found that the change in upper critical field (magnetic field required for AFM to FM transition) depends sensitively on the grease. Since critical field is higher for sample mixed with silicon grease, the effect of grease appears to be equivalent to pressure. 

\begin{figure}[t]
	\begin{center}
	\includegraphics[width=8 cm] {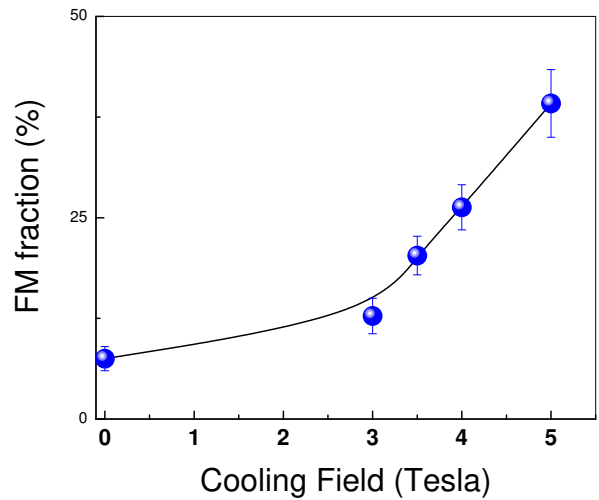}
	\end{center}
	\caption{(color online) Variation of ferromagnetic (FM) fraction in remanant state at 5 K as a function of cooling magnetic field  obtained from the analysis of $^{57}Fe$ M$\ddot{o}$ssbauer spectra shown in figure 2.  The solid line is a guide to eye.}
	\label{Figure3}
\end{figure}

\begin{figure}[b]
	\begin{center}
	\includegraphics[width=8 cm] {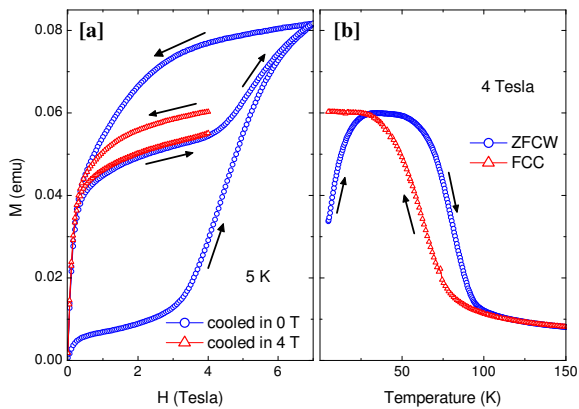}
	\end{center}
	\caption{(color online) \textbf{[a]} Magnetization (M-H) of $Hf_{0.77}Ta_{0.23}Fe_2$ powder mixed with silicone grease (as used in M$\ddot{o}$ssbauer experiemnts). Curve represented by open circle (blue) measured after cooling in zero field and magnetic field is varied fro 0 to 7 to 0 to 7 Tesla. Curve represented by red triange measured after cooling in 4 Tesla from 150 K and magnetic field is varied from 4 to 0 to 4 Tesla. \textbf{[b]} Temperature dependence of magnetization measured in the presence of 4 Tesla during warming after zero field cooling (ZFCW) and subsequent cooling (FCC). Arrows indicate temperature and magnetic field sweep directions.}
	\label{Figure4}
\end{figure}

It is to be noted that when the external magnetic field is applied at 5K (i.e., when the sample is cooled to 5K in ZFC and the external magnetic field is applied), no enhancement of FM fraction is observed (figure 6(a)) from M$\ddot{o}$ssbauer measurements.  It shows that the fraction of FM and AFM states at 5K depends on the path followed in H-T space to reach the measurement temperature. 

Figure-5(a) shows the M$\ddot{o}$ssbauer spectrum measured at 150K with external magnetic field (4 Tesla). At 150K, the sample is in complete AFM state.  The in-field M$\ddot{o}$ssbauer data at 150K is fitted with two sites corresponding to AFM(6h) and AFM(2a). The evolution of M$\ddot{o}$ssbauer spectrum with the application of magnetic field depends on the nature of magnetic ordering present in the samples \cite{Chap1979}.  The width of the lines, separation of outer lines and the $A_{23}$ give unambiguous information about the magnetic texture present in the compound. When the M$\ddot{o}$ssbauer spectrum of AFM sample is measured under the application of high magnetic fields, various phenomena, namely, spin-flop, spin-canting and parallel spin alignment can happen \cite{Chap1979}, depending on the interplay between the $B_{INT}$, anisotropy field ($B_A$) and $B_{EXT}$.  All these transitions can be clearly observed in single crystal AFM samples and often the observed lines are sharp. However, in polycrystalline AFM samples because the angle between the $B_{EXT}$ and the easy axis is evenly distributed between 0 and $\pi$, the experimental signal is then a superposition of all the constituent crystallite signals, and therefore one observes a broadened line with an effective hyperfine field $B_{EFF}=(B^2_{INT}+B^2_{EXT})^{1/2}$.  And the intensity of the lines corresponding to the $\Delta m=0$ transition gives the information about the spin configuration, which is usually estimated by measuring $A_23$ parameter.  Whereas, for a non-magnetic sample one would observe a hyperfine splitting equal to that of $B_{EXT}$. However, if any moment is induced with the application of magnetic field then the observed $B_{EFF}$ would not be exactly equal to $B_{EXT}$. Since the strength of the hyperfine field and the quadrupole splitting values corresponding to AFM(2a) site are comparable, the data is fitted with the following Hamiltonian \cite{Gree1971} corresponding to this site

\

$H = -g \mu_N \hat{I}_z H + \frac{e^2qQ}{4I(2I-1}[3I^2_z-I(I+1)] $

\

where, the first part is the magnetic interaction and the second part is the quadrupole interaction.  

\begin{figure}[htb]
	\begin{center}
	\includegraphics[width=8 cm] {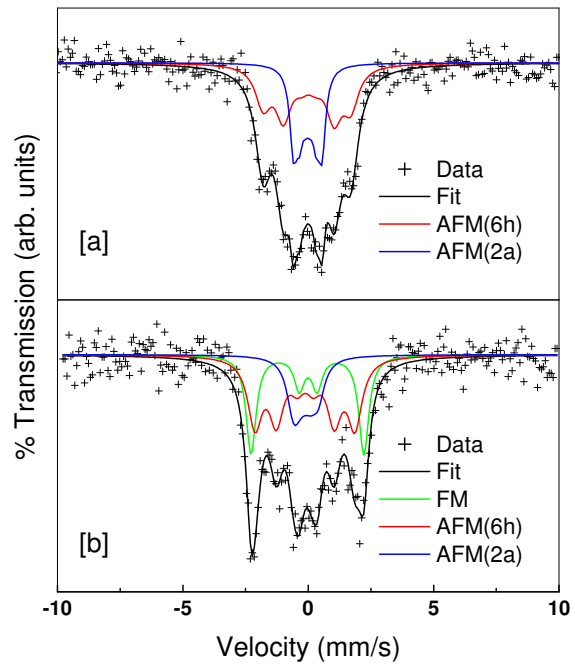}
	\end{center}
	\caption{(color online) High field (4 Tesla) $^{57}Fe$ M$\ddot{o}$ssbauer spectra of $Hf_{0.77}Ta_{0.23}Fe_2$ measured at  (a) 150K and (b) 5K.  The dots are the experimental points and the solid lines are the best fits to the data. Obtained hyperfine parameters are shown in Table-III.}
	\label{Figure5}
\end{figure}

	The area ratio of these two sites i.e., AFM(6h) and AFM(2a)  is constrained to be in the ratio of 3:1 respectively (this is adopted for all the high field data in the present work) and all other parameters were kept free. The obtained hyperfine parameters are shown in Table-III.  The spectrum corresponding to the AFM(6h) site exhibits a broadened sextet with an $B_{EFF}$ of about 10.84 Tesla.  Whereas, for the AFM(2a) site, one would expect a $B_{EFF}$ equal to the $B_{EXT}$, but the observed $B_{EFF}$ is less than $B_{EXT}$ indicating that there is a finite moment generated on this site, which is parallel to the $B_{EXT}$.   The important point to be noted here is the value of $A_{23}$ i.e., the area ratio of the second and third line intensities corresponding to the AFM(6h) site.  This parameter gives information about the angle ($\theta$) between the $\gamma$-rays (parallel to $B_{EXT}$ in the present case) and the $B_{EFF}$ and is estimated using the equation $cos^2\theta = (4-A_{23})/(4+A_{23})$  \cite{Chap1979}.  The angle comes out to be about $82^0$ (at 150K and 4 Tesla) which indicates spin-flop transition, in which the moments are aligned nearly perpendicular to the $B_{EXT}$.  
	
\begin{table} [b]
\caption{\label{arttype}Obtained hyperfine parameters from the fitting of $^{57}Fe$ M$\ddot{o}$ssbauer spectra measured in the presence of 4 Tesla external magnetic field at the indicated temperatures (shown in figure 5). FCC indicates that measurement temperature is reached by cooling in the presence of 4 Tesla magnetic field.}
\begin{ruledtabular}
\begin{tabular}{cccccc}

Temp& FWHM & $B_{EFF}$	& $A_{23}$	& Area & Phase\\
   & (mm/s) & (Tesla)	&  & \% & \\

150 K     & 0.741$\pm$0.034 & 10.84$\pm$0.11	& 3.84$\pm$.27	& 75.2 & AFM(6h)\\
          & 0.352$\pm$0.032 & 3.53$\pm$0.08   & -	            & 24.8 & AFM(2a)\\
5 K  & 0.459$\pm$0.063 & 13.81$\pm$0.14	& 0.00	        & 24.2 & FM    \\
(FCC)         & 0.791$\pm$0.070 & 12.12$\pm$0.36	& 2.67$\pm$0.60	& 57.0 & AFM(6h)\\
          & 0.589$\pm$0.093 & 3.17$\pm$0.22	  & -	            & 18.8 & AFM(2a)\\

\end{tabular}
\end{ruledtabular}
\end{table}

Figure-5(b) shows the M$\ddot{o}$ssbauer spectrum measured at 5K with external magnetic field (4 Tesla).  The temperature 5K was approached in field cooled cooling (FCC) condition from 150K.  The high field data of 5K is fitted with a convolution of three sub-spectra corresponding to the FM and AFM (6h and 2a sites).  For a FM sample, what is observed with the application of $B_{EXT}$ is the effective hyperfine field ($B_{EFF}$) defined as $\vec{B}_{EFF}= \vec{B}_{INT}+\vec{B}_{EXT}$ (neglecting the demagnetizing fields) would either decrease or increase depending on whether the magnetic moment is anti-parallel/parallel to the $B_{INT}$.  Since the FM component is expected to be magnetically soft and aligns with the $B_{EXT}$, the $A_{23}$ parameter of this sextet is fixed as zero.  The obtained hyperfine parameters are shown in Table-III.  As one can see, the hyperfine field of FM component is reduced with the application of external field indicating that the moment is anti-parallel to the $B_{INT}$, similar to most of the compounds \cite{Chap1979}.  A magnetic moment ($\mu_{FM}$) of about 0.35 $\mu_B$ is calculated due to FM component from the observed $B_{EFF}$ using the equation $B_{EFF}= A\mu_B$, where A is the constant, whose value is taken as 9.5 Tesla/$\mu_B$ from the study of Beloscvic et al. \cite{Belo2006}, on this compound.  The spectrum corresponding to AFM phase is similar to that of 150K (figure 5(a)) data as discussed above.  But the observed value of $A_{23}$ parameter corresponding to AFM(6h) site  is drastically different from the 150K data and the angle ($\theta$) between the external magnetic field ($\gamma$-ray) and the Fe magnetic moment comes out to be about $64^0$, which indicates that the AFM(6h) site moments are canted with respect to the applied magnetic field.  As a result of spin-canting one observes a larger magnetization arising due to the AFM phase ($\mu_{AFM}$), which is calculated as the cosine component in the direction of $B_{EXT}$ and has a value of about 0.32 $\mu_B$.  Therefore, from the present high field M$\ddot{o}$ssbauer data, we conclude that the contribution of AFM component at 4 Tesla and 5K to the total magnetization is almost comparable to that of FM component contribution. The observed increase in the canting of AFM spins at 5K as compared to 150K indicates that there is some sort of magnetic interaction between the FM and AFM components.

\begin{figure}[htb]
	\begin{center}
	\includegraphics[width=8 cm] {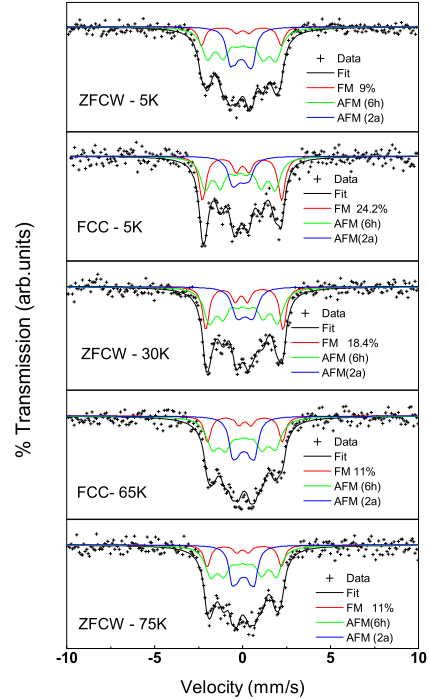}
	\end{center}
	\caption{(color online) High field (4 Tesla) $^{57}Fe$ M$\ddot{o}$ssbauer spectra of $Hf_{0.77}Ta_{0.23}Fe_2$ measured at temperatures approached in field cooling (FCC) and zero-field cooled warming (ZFCW) 150K and 5K.  The dots are the experimental points and the solid lines are the best fits to the data.}
	\label{Figure6}
\end{figure}

\begin{figure}[htb]
	\begin{center}
	\includegraphics[width=8 cm] {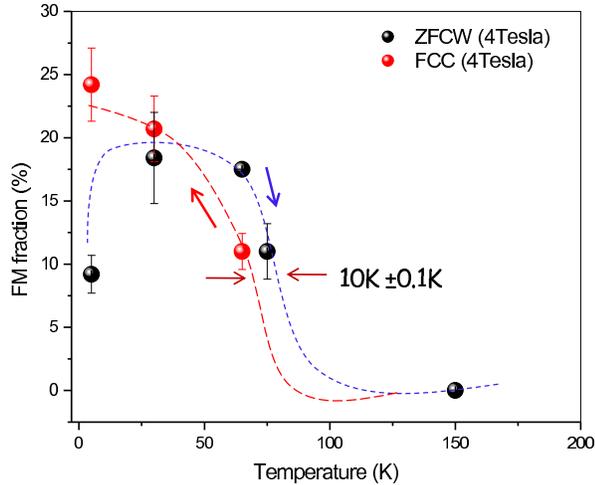}
	\end{center}
	\caption{(color online) Variation of arrested ferromagnetic (FM) fraction as a function of temperature obtained in FCC and ZFCW protocols.  The dotted line is a guide to eye.}
	\label{Figure7}
\end{figure}

To determine the equilibrium state of the system we carried out the M$\ddot{o}$ssbauer measurement under CHUF protocol. The high field (with an application of 4 Tesla field parallel to the gamma rays) M$\ddot{o}$ssbauer measurements are carried out at different temperatures under two  protocols, namely, field-cooled cooling (FCC) and zero-field cooled warming (ZFCW).  For FCC measurements, the $B_{EXT}$ is applied and the sample is cooled from 150 K to the temperature of interest in the presence of magnetic field.  Whereas in ZFCW protocol, the sample was cooled from 150K to 5K in zero field and 4 Tesla magnetic field is applied at 5K and the sample is heated to temperature of interest in the presence of magnetic field.  In both FCC and ZFCW protocols, the M$\ddot{o}$ssbauer data is recorded in the presence of applied magnetic field.  Figure 6 shows the representative high field M$\ddot{o}$ssbauer data measured at 4 Tesla and different temperatures approached in different protocols, namely, FCC and ZFCW.  The fits to the data were carried out in a similar manner to that as discussed above to extract the fraction of FM component. Thus, the obtained FM fraction from this analysis is plotted in figure 7 as a function of temperature.  For ZFCW in the presence of 4 Tesla, the FM fraction increases to 17\% on warming from 5 K to 30 K. With further increase in temperature, the FM fraction starts decreasing above 65 K, therefore, showing a reentrant transition. It shows that FM state is the equilibrium state at 5K and 4 Tesla consistent with bulk magnetic measurement. During FCC in the presence of 4 Tesla, it shows AFM to FM transition with sharp increase in FM fraction around 65 K. The FM fraction at 65 K during FCC is found to be identical to that observed at 75 K during ZFCW curve, thereby giving a hysteresis width of about 10 K of the first order AFM-FM transition. Comparatively, smaller hysteresis width could be associated with much longer measurement time (~70 hours) for M$\ddot{o}$ssbauer measurements as compared to magnetization measurements which are performed at the rate of 1.5 K/min.

\maketitle\section{Conclusions}
In conclusion, low temperature high magnetic field (LTHM) $^{57}Fe$ M$\ddot{o}$ssbauer measurements were carried out on inter-metallic $Hf_{0.77}Ta_{0.23}Fe_2$ compound to demonstrate the phenomena of kinetic arrest and de-vitrification, which are recently reported in this compound using bulk magnetization measurements. It is also observed that, in the presence of a ferromagnetic component, the antiferromagnetic spins cant with respect to the applied magnetic field and hence, contribute to the total bulk magnetization in this compound.

\maketitle\section{Acknowledgements}
Cryogenics, CSR Indore is acknowledged for providing liquid helium for LTHM M$\ddot{o}$ssbauer experiments. Dr. R J Chaudhari is acknowledged for magnetization measurements.

\end{document}